\documentclass[12pt]{article}
\usepackage{graphicx,psfrag,epsf}
\usepackage{caption}
\usepackage{hyperref}
\usepackage{enumerate}
\usepackage{natbib}
\usepackage{url} 
\usepackage{hhline,array,amsmath,graphicx}
\usepackage{pifont,latexsym,ifthen,rotating,calc}
\usepackage{siunitx}
\usepackage{etoolbox}
\usepackage{mathtools}
\usepackage{amsmath}
\usepackage{amssymb}
\usepackage{amsthm}
\usepackage{bbm}
\usepackage{pdflscape}
\usepackage{rotating}
\usepackage{bm}
\usepackage{enumitem}
\usepackage[mathscr]{euscript}
\usepackage{multirow}
\usepackage{xcolor}
\usepackage{booktabs}
\usepackage[margin=0.85in]{geometry}

\newcommand{\blind}{0}

\begin{document}

\def\spacingset#1{\renewcommand{\baselinestretch}%
{#1}\small\normalsize} \spacingset{1}

\date{}


\if0\blind
{
  \title{\bf Bayesian Multiple Testing for Suicide Risk in Pharmacoepidemiology: Leveraging Co-Prescription Patterns}
  \author{Soumya Sahu\\
    Department of Epidemiology and Biostatistics, University of Illinois Chicago\\
    Kwan Hur\\
    Center for Health Statistics, The University of Chicago\\
    Dulal K. Bhaumik\\
    Department of Epidemiology and Biostatistics,\\
    Department of Psychiatry,
    University of Illinois Chicago\\
    Robert Gibbons\\
    Center for Health Statistics, The University of Chicago\\
    }
  \maketitle
} \fi

\if1\blind
{
  \bigskip
  \bigskip
  \bigskip
  \begin{center}
    {\Large \bf Bayesian Multiple Testing for Suicide Risk in Pharmacoepidemiology: Leveraging Co-Prescription Patterns}
\end{center}
  \medskip
} \fi

\bigskip

\begin{abstract}
Suicide is the tenth leading cause of death in the United States, yet evidence on medication-related risk or protection remains limited. Most post-marketing studies examine one drug class at a time or rely on empirical-Bayes shrinkage with conservative multiplicity corrections, sacrificing power to detect clinically meaningful signals. We introduce a unified Bayesian spike-and-slab framework that advances both applied suicide research and statistical methodology. Substantively, we screen 922 prescription drugs across 150 million patients of U.S. commercial claims (2003–2014), leveraging real-world co-prescription patterns to inform a covariance prior that adaptively borrows strength among pharmacologically related agents. Statistically, the model couples this structured prior with Bayesian false-discovery-rate control, illustrating how network-guided variable selection can improve rare-event surveillance in high dimensions. Relative to the seminal empirical-Bayes analysis of Gibbons et al. (2019), our approach reconfirms the key harmful (e.g., alprazolam, hydrocodone) and protective (e.g., mirtazapine, folic acid) signals while revealing additional associations, such as a high-risk opioid combination and several folate-linked agents with potential preventive benefit that had been overlooked. A focused re-analysis of 18 antidepressants shows how alternative co-prescription metrics modulate effect estimates, shedding light on competitive versus complementary prescribing. These findings generate actionable hypotheses for clinicians and regulators and showcase the value of structured Bayesian modelling in pharmacovigilance.
\end{abstract}

\noindent%
\vfill

\newpage
\spacingset{1.8} 
\section{Introduction}

\subsection{Problem of Interest} 
Suicide is a major public health crisis, consistently ranked among the top causes of mortality in the United States. According to \cite{Bachmann2018}, in 2018 alone, it claimed more than 48,000 lives, making it the tenth leading cause of death (and the second among young adults).  Alarmingly, US suicide rates rose by roughly 30\% from 2000 to 2018.  Identifying medication-related suicide risk (or protection) is exceptionally challenging. Most suicides occur alongside psychiatric disorders, making it difficult to isolate drug effects due to confounding by illness severity and other factors. Clinical trials are typically underpowered to detect such rare events, and post-market evidence is often observational and prone to bias.  

Regulatory agencies have long grappled with this uncertainty. The US Food and Drug Administration (FDA) has issued ``black box'' warnings for more than 130 medications on potential suicide-related adverse effects (\cite{Lavigne2012}). For example, in 2004 the FDA warned that antidepressants might increase suicidal ideation in youth; in 2008 a class-wide warning was added for antiepileptic drugs. These actions can have profound public health implications: after the 2004 warning, antidepressant prescriptions for young people dropped, and suicide attempt rates appeared to increase in that population (\cite{Lu2014}). Such cases underscore the high stakes and the need for solid evidence. Currently, clinicians and patients are often left to weigh the benefits of medications against the uncertain risks of suicide (\cite{Lavigne2012}). Hundreds of millions of prescriptions are filled annually for drugs carrying suicide-risk warnings (\cite{Lavigne2012}), yet for most, there is little quantitative data on how much (if at all) they actually modulate the risk of suicide. This highlights the urgent need for robust, data-driven tools to detect harmful or protective drug effects on suicidal behavior in real-world data.

Recent large-scale pharmacoepidemiologic studies illustrate the promise of data-driven approaches. In particular, \cite{Gibbons2019} analyzed medical claims for 922 prescription drugs (covering more than 150 million patients) to screen for associations with suicide attempts and self-harm. Using a novel high-dimensional empirical Bayes algorithm, they discovered 10 drugs associated with significantly \emph{increased} suicide attempt rates and 44 drugs associated with \emph{decreased} rates. Among the strongest risk signals were certain sedatives and analgesics (e.g., alprazolam and opioid combinations), while unexpectedly protective signals included medications such as folic acid and the antidepressant mirtazapine.  Interestingly, many drugs flagged as potentially protective had previously been under suspicion – in fact, 30 of the 44 protective signals were for psychotropic drugs (antidepressants, antipsychotics, etc.) that carry FDA suicide warnings.  The standout finding was that patients filling folic acid (vitamin B$_9$) prescriptions had substantially lower rates of suicide attempts, a signal later confirmed in a dedicated follow-up study in \cite{Gibbons2022} showing about a 40–45\% reduction in suicidal events associated with folate supplementation.  Such discoveries are highly motivating: they hint that improved surveillance can not only identify drugs that increase risk, but also reveal overlooked preventive opportunities (folic acid being a safe, inexpensive supplement that could potentially be repurposed to mitigate suicide risk). The U.S. Department of Defense has recently funded a large-scale randomized clinical trial to confirm this association. If confirmed, it will provide an unprecedented advance in suicide prevention.

Importantly, patients are often concurrently prescribed multiple medications, either for comorbid conditions or as part of combination therapy (\cite{Bazzoni2015}). These co-prescription patterns introduce dependencies among drug exposures, which, if ignored, can lead to biased or inefficient inference. Robust analytical frameworks must account for associations among drugs to accurately detect true risk or protective effects. Given the public health burden of suicide and gaps in current knowledge, more powerful statistical tools are needed to mine high-dimensional healthcare data for medication safety signals.

\subsection{Relevant Statistical Tools and Literature} 

Our problem lies at the intersection of high-dimensional multiple testing and observational drug safety analysis. Analyzing hundreds of drugs raises a severe multiple comparisons challenge. Classical corrections such as Bonferroni \cite{Bonferroni1936} or Benjamini--Hochberg (BH) \cite{Benjamini1995} are often too conservative or ill-suited for rare, correlated events. In pharmacoepidemiology, empirical Bayes (EB) methods—such as the shrinkage model by \cite{DuMouchel1999}, used in the FDA’s reporting system—offer strength-sharing across tests via data-driven priors. However, despite their computational appeal, EB methods face key limitations in sparse, high-dimensional drug safety settings.

\cite{Gibbons2019} adopted the empirical Bayes framework developed in \cite{Hedeker2006}, applying a mixed-effects logistic regression model in which drug-specific slopes were treated as random effects drawn from a normal distribution with unknown variance. This prior variance was estimated using empirical mean squared error across drugs, leading to a classic marginal maximum likelihood EB estimator. While elegant, this approach introduces several challenges in sparse event settings. First, since suicide events are rare, the marginal likelihood often favors a small prior variance estimate, which results in excessive shrinkage of all drug-specific effects toward zero. As discussed in \cite{Carlin2000}, this can lead to underestimation of uncertainty and systematic failure to detect weak but real signals.
Second, the EB approach used by \cite{Gibbons2019} assumes that the posterior distribution of each drug effect is approximately normal, and computes z-scores from estimated posterior means and variances. In rare-event settings, this normality assumption often fails, especially when the number of events per drug is very low. The consequence is that uncertainty may be poorly quantified, resulting in confidence intervals that are misleadingly narrow or p-values that are miscalibrated (\cite{Carlin2000, ScottBerger2010}). Third, the EB framework does not explicitly model the null hypothesis. All drug effects are treated as random draws from a single continuous distribution centered at zero, with no point mass at zero to reflect true null effects. This structure precludes the calculation of posterior probabilities of null vs. non-null status, complicating interpretation of p-values derived from approximate posteriors. As noted by \cite{ScottBerger2010}, this lack of explicit null modeling weakens the foundation for formal decision-making in hypothesis testing. Furthermore, \cite{Cui2008} argues that the EB assumption of fixed prior parameters, estimated from the data, can result in biased selection decisions in sparse signal regimes. Finally, \cite{Gibbons2019} analysis applies a Bonferroni-style multiple testing correction to the EB-derived z-scores and p-values, which is known to be highly conservative—particularly problematic in high-dimensional settings with subtle effects. This compounding of conservative correction with over-shrinkage reduces the power to detect important safety signals, especially those with modest effects.
Together, these limitations illustrate that empirical Bayes methods, while computationally tractable, may be ill-suited to the demands of rare-event drug safety surveillance. In contrast, fully Bayesian approaches offer a compelling alternative: by treating all parameters, including hyperparameters and sparsity levels, as random variables with appropriate priors, they allow for full posterior uncertainty propagation and enable rigorous inference on both effect sizes and null probabilities. Thus, such a framework can flexibly model complex dependence, adapt to sparse data, and deliver more principled signal detection in high-dimensional pharmacoepidemiologic settings.

Bayesian spike-and-slab priors provide a natural framework for sparse signal detection. Originally introduced by \cite{Mitchell1988} and extended by George and McCulloch (1993), spike-and-slab priors assume each drug’s effect (say $\theta_j$) is either exactly zero (“spike” at 0) or drawn from a broad distribution (“slab”) if the drug is truly associated with the outcome. This two-component mixture prior encourages drug effects to be null while allowing deviation if drugs have an impact on suicide risk. Posterior inference under spike-and-slab yields a posterior inclusion probability (PIP) for each drug – essentially, the probability that drug $j$ is truly associated with the outcome (i.e. $\theta_j$ in the slab). These PIPs serve as a Bayesian analog of ``significance" measures for each hypothesis. Building on the notion of PIPs, researchers have developed principled ways to control the false discovery rate (FDR) in a Bayesian manner. Rather than $p$-values and $q$-values, one can threshold on PIPs to decide which drugs to report, aiming to limit the expected proportion of false positives among those selected. \cite{Scott2006} showed that one can achieve adaptive multiplicity control by using a Beta-Bernoulli prior on model inclusion – effectively making the cutoff for ``significance" more stringent as the number of tests increases. \cite{Muller2007} took a decision-theoretic approach and derived an optimal Bayesian rule controlling FDR under a specific loss function. Their rule selects hypotheses in order of descending PIP (or posterior odds) until a certain posterior expected false discovery cost is met, yielding a procedure analogous to the classical BH method but in the posterior probability domain. In practice, this means we can choose a PIP threshold $t$ such that the Bayesian FDR (the expected false discovery proportion conditional on the observed data) is below a target level (e.g. 10\%). Such Bayesian FDR-controlling procedures have been successfully applied in genomics and other high-dimensional settings, and they provide a rigorous yet convenient way to call out likely signals while managing the error rate. We adopt a similar approach here, using PIPs as our primary metric of evidence and selecting drug signals via Bayesian FDR control.

An important challenge in high-dimensional screening is how to incorporate structure or prior knowledge across hypotheses. In pharmacovigilance, patients often take multiple medications, inducing correlations in drug exposures. Co-prescription patterns—reflecting shared indications or patient profiles—can be represented as networks, offering a data-driven way to encode drug similarity. Incorporating such structure into statistical models via network-based regularization or covariance priors enables borrowing of strength across related drugs \cite{Bazzoni2015}. However, leveraging co-prescription information within high-dimensional Bayesian selection remains relatively underexplored in drug safety research.

\subsection{Our Contributions} 
This paper develops a new Bayesian modeling framework for detecting drug-associated suicide risk or any rare adverse event, addressing the challenges outlined above. Ultimately it is a replacement for pharmacovigilance signal generation methods currently used by the US FDA based on spontaneously reported adverse events (see \cite{gibbons2015statistical}). We advance the state of the art in several important ways:

(1) Fully Bayesian spike-and-slab selection:  We replace the empirical Bayes paradigm of prior work \cite{Gibbons2019} with a fully Bayesian hierarchical model. Each drug’s effect on the suicide outcome is given a spike-and-slab prior, allowing most drugs to have essentially zero effect while a few have nonzero effects. Crucially, we treat all hyperparameters (such as the prevalence of risk signals and the effect size distribution) as unknown and assign them proper priors, rather than estimating them via empirical Bayes. Markov Chain Monte Carlo (MCMC) is used to sample from the joint posterior. This fully Bayesian treatment properly accounts for uncertainty in the hyperparameters and adaptively tunes the amount of shrinkage to the data, which is particularly important in our extremely sparse signal setting. As a result, our model is more sensitive in detecting true signals than the previous empirical Bayes screening approach, which tended to over-shrink estimates when events were rare.

(2) Incorporation of co-prescription network information: We introduce a novel prior covariance structure on the drug effects that leverages co-prescription patterns. We construct a drug-drug co-prescription matrix from real-world prescription data, where entries reflect how often two drugs are used by the same patients (after adjusting for overall usage rates). This matrix is used to inform a covariance matrix in the prior for $\boldsymbol{\theta} = (\theta_1,\dots,\theta_p)$. In essence, our model assumes that if two drugs are commonly co-prescribed, their log odds ratios for suicide are \emph{a priori} likely to be similar. We implement this through a multivariate spike-and-slab prior: the ``slab" component is a multivariate Normal distribution on the vector of effects, with a covariance matrix proportional to the co-prescription affinity matrix. To our knowledge, this is the first pharmacovigilance application of a network-informed prior on drug effects. By borrowing strength across related drugs through the co-prescription matrix, the model can more readily detect weak signals for a given drug when its co-prescribed counterparts exhibit stronger evidence. This structured prior promotes coherent detection across related drugs, improving sensitivity to subtle effects without inflating false discoveries.

(3) Posterior inference and Bayesian FDR control for signal discovery: From the MCMC output, we obtain posterior inclusion probabilities for each drug – the probability that drug $j$ is associated with increased or decreased risk. We then employ a Bayesian FDR-controlling procedure \cite{Scott2006, Muller2007} to decide which drugs to label as discoveries. Specifically, we rank drugs by their PIP and select a threshold $t$ such that the expected false discovery proportion is below a desired level (e.g. 5\%). This means that if, say, 20 drugs are flagged, we expect, on average, only 1 of them to be a false lead at $5\%$ FDR. This rigorous approach to identifying signals replaces the heuristic p-value thresholds or ad-hoc rule used in earlier work. It provides a principled way to balance the discovery of true risk markers against the rate of false alarms. Notably, our Bayesian FDR approach naturally handles two-sided signals (increased and decreased risk) by computing each drug’s probability of having any effect (harmful or protective) and controlling the false discovery rate for the two-directional discovery set.

We validate our methodology through extensive simulation studies designed to closely mimic the high-dimensional, rare-event characteristics inherent to the suicide risk assessment problem. Specifically, our simulations considered scenarios involving both large-scale settings (922 drugs, sparse true signals) and moderate-scale settings (100 drugs, fewer signals). The simulation results clearly demonstrate that our fully Bayesian spike-and-slab model, particularly when leveraging co-prescription-informed priors, achieves substantially higher true positive detection rates while simultaneously maintaining precise control over false discoveries.

Applying our method to the original extensive medical claims dataset previously analyzed by \cite{Gibbons2019}, we not only successfully confirmed the majority of the original findings but also uncovered numerous additional drug signals missed by earlier empirical Bayes analyses. Specifically, our approach identified several drugs associated with reduced suicide risk, including agents closely related to folic acid and its metabolic pathway, further reinforcing previous observations on folate supplementation's potential preventive benefits. Moreover, we identified new high-risk signals for several antidepressants whose association appeared marginal under previous methods but became clear when incorporating co-prescription dependencies.


\noindent \textbf{Non-Causal Interpretation of Findings.} Our goal is signal detection of associations, not causal effect estimation. Our approach compares the rate of the adverse event before and after exposure in the same individuals, eliminating between-person confounding, and provides a proper estimate of the rate of the event before and after. However, time-varying confounding (e.g., evolving illness severity, concomitant care) may still influence associations. Accordingly, discoveries should be viewed as hypothesis-generating and candidates for targeted causal follow-up (e.g., target trial emulation, self-controlled case series variants, instrumental variable analyses).

\noindent \textbf{Novelty of Our Approach.} We emphasize that some building blocks of our framework are standard in the Bayesian literature (e.g., spike–and–slab priors, posterior inclusion probabilities, and Bayesian FDR control). The novelty of our work lies in how these elements are integrated, specialized, and validated for pharmacovigilance of suicide risk at the national scale. Specifically: (i) we introduce a co-prescription–informed covariance on drug-specific random slopes within a within-person incident user cohort design, allowing principled information sharing across drugs that are jointly used in practice (including attenuation of borrowing for negatively associated, substitutable drugs); (ii) we operationalize this via a multivariate spike–and–slab prior on the vector of drug effects with a Kronecker structure that couples a learnable within-drug covariance to a data-derived co-prescription matrix; and (iii) we provide a decision-theoretic, two-sided Bayesian FDR procedure tailored to rare-event screening, including practical calibration of the control parameter under extreme sparsity. Beyond methodology, the paper presents a scalable analysis of 922 drugs in U.S. claims data that reproduces known signals and identifies clinically credible new ones, demonstrating that a structured, co-prescription-aware Bayesian screening can enhance power without inflating false discoveries. In this sense, our contribution is a problem-driven synthesis that advances both the applied suicide-risk literature and the statistical practice of high-dimensional safety surveillance.

\section{Data Description}

\textbf{Claims Data and Suicide Outcomes.}
The primary data source for this study is a large administrative health claims database covering commercially insured individuals in the United States from 2003 to 2014, previously analyzed in  \cite{Gibbons2019}. This dataset comprises over 150 million unique patients and includes detailed, time-stamped records of filled prescriptions, outpatient and inpatient encounters, and diagnostic codes. Each entry contains information on drug dispensing, prescriber identifiers, dates of service, patient demographics (age, sex), and International Classification of Diseases (ICD) diagnosis codes.

For the purpose of studying suicide risk associated with medication exposure, the analysis focused on a curated list of 922 prescription drugs spanning diverse therapeutic classes. This includes all drugs on the market that had more than 3,000 prescription in 2014. The outcome of interest was a suicide attempt or intentional self-harm event, identified using ICD-9 and ICD-10 diagnostic codes consistent with definitions from the Centers for Disease Control and Prevention (CDC). These codes encompass a range of suicidal behaviors, from non-fatal self-poisoning to violent self-injury and completed suicides, though the data largely reflect attempts that resulted in clinical encounters.

\noindent \textbf{Study Design and Data Structure.} 
We follow the within-person incident user cohort (WPIUC) design used by \citet{Gibbons2019}. For each drug, we define a cohort of \emph{new users}: patients with their first observed dispensing of that medication during the study period (2003-2014). For each eligible patient–drug pair, we define two equal-length observation windows around the dispensing date: a \emph{pre-exposure} window (the 90 days before dispensing) and a \emph{post-exposure} window (the 90 days after dispensing). Events occurring on the dispensing date are excluded, and sensitivity analyses can further exclude events in the first week after dispensing. This construction creates matched, within-person strata that compare post- versus pre-exposure risk for the same individual, thereby controlling for all time-invariant confounders (baseline psychiatric history, genetic predisposition, socioeconomic status). 

The analysis dataset aggregates counts within strata defined by drug, a small set of demographic strata (age, sex, and their interaction), and time (pre vs.\ post). For each stratum we record the number of persons at risk and the number of suicide-related events. This stratification allows us to estimate drug-specific exposure effects while characterizing how associations may differ across age–sex strata. An example for the first nine drugs (with columns: Drug ID, Age, Sex, Age$\times$Sex, Time, $N$, and number of suicide attempts) is provided in Supplementary Table 4, which illustrates the exact structure used in estimation.

\noindent \textbf{Co-Prescription Information.}
In addition to the historical claims data used for modeling drug–suicide associations, we incorporated external co-prescription information to characterize how antidepressant medications are used in real-world clinical practice. Specifically, we obtained co-prescription frequency tables for the calendar years 2020 and 2021, each derived from aggregated pharmacy records covering a broad population of patients. These data capture the number of individuals who were prescribed any pair of antidepressants within the same calendar year, thus reflecting patterns of concurrent medication use.

\begin{figure}
    \centering
    \includegraphics[width=0.7\linewidth]{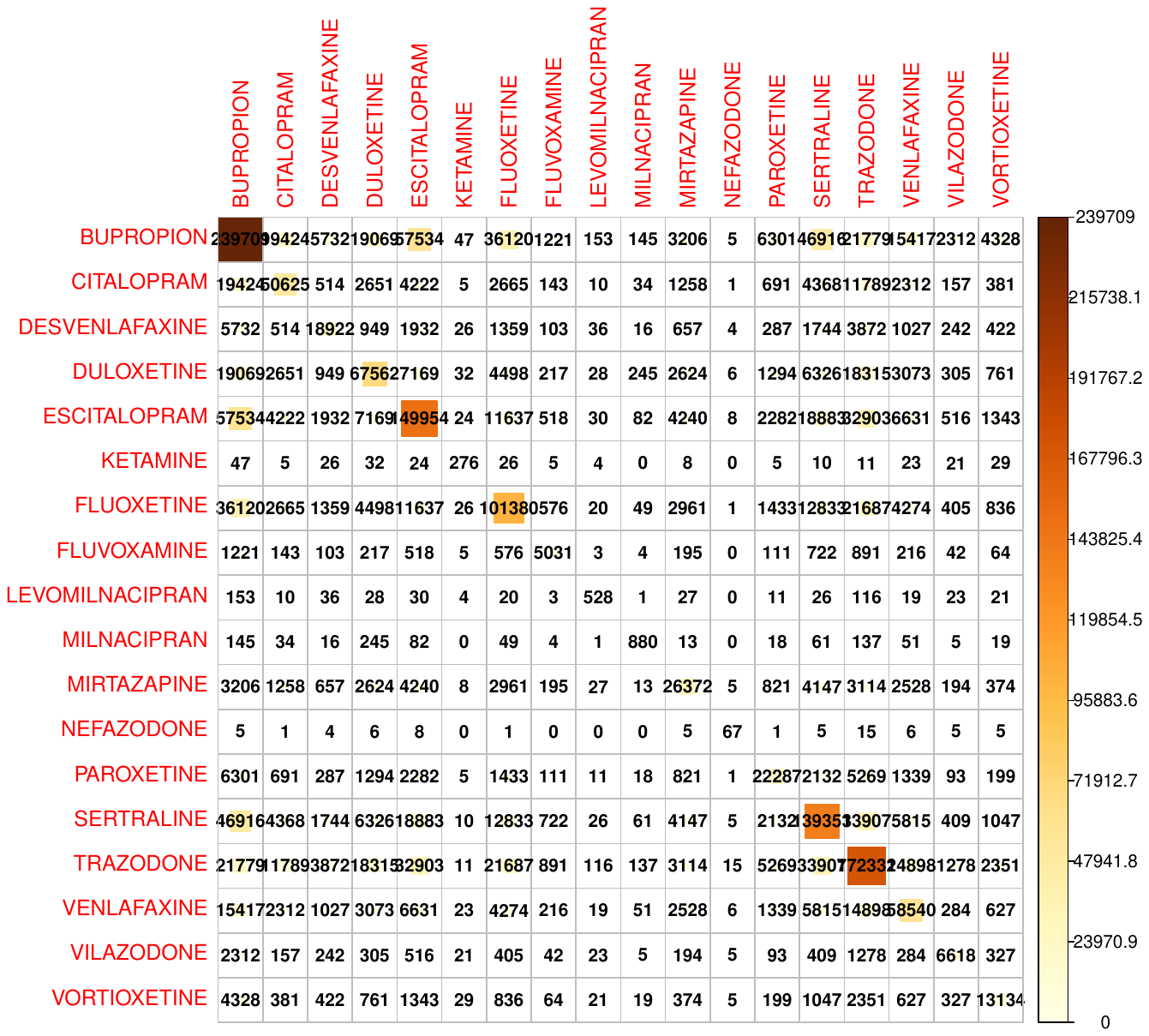}
    \caption{\footnotesize Co-prescription table for 18 antidepressant drugs for Calender year 2021: Rows and columns correspond to individual antidepressant drugs, and entries represent the number of patients who filled prescriptions for both drugs}
    \label{fig:co_pres_num}
\end{figure}

Each year's data can be organized as a symmetric square table (see figure \ref{fig:co_pres_num}), where rows and columns correspond to individual antidepressant drugs, and entries represent the number of patients who filled prescriptions for both drugs. These tables include both high-frequency combinations (e.g., SSRIs with atypical antidepressants) and less common co-prescriptions. Since patients often receive multiple antidepressants simultaneously to manage complex or treatment-resistant mood disorders, these data provide a useful empirical snapshot of polypharmacy behavior in psychiatric prescribing.

While these co-prescription data postdate the main claims dataset used for outcome modeling, they serve as an important auxiliary source for informing the prior structure in our Bayesian model. Specifically, these frequencies reveal which drugs are most likely to be used in tandem, offering a clinically meaningful proxy for latent relationships among medications. This information allows us to define similarity or proximity among drugs based on usage context, independent of their chemical structure or official therapeutic class. By incorporating co-prescription data into our statistical framework, we are able to introduce structure into the prior distribution of drug effects—enabling more coherent signal detection among pharmacologically or clinically related drugs. The integration of co-prescription information into the modeling process is described in detail in Section 3, where we show how these pairwise frequencies are transformed into a co-prescription matrix and used to inform a structured prior covariance on drug-specific effects.

\section{Model}\label{sec:model}
Our modeling framework is built upon the WPIUC design described in \cite{Gibbons2019}, which allows each subject to serve as their own control by comparing suicide-related event rates before and after drug exposure. This design facilitates the estimation of within-subject causal effects while implicitly adjusting for all time-invariant confounding. The statistical validity of the estimator relies on several key assumptions, which we describe and explain below.

Let \( A_i \) be the event that an individual experiences an adverse event (AE) of interest related to drug \( i \), and let \( T_{A_i} \) denote the time at which this event occurs. In the case of recurring events, \( T_{A_i} \) may be a vector, but for the purposes of this study, we focus on the first occurrence within a predefined window. Similarly, let \( D_i \) denote the event that the individual fills a prescription for drug \( i \), and \( T_{D_i} \) be the time of the drug fill. We define a binary exposure indicator \( x_i \) such that:
\[
x_i = 
\begin{cases}
0, & \text{if } T_{A_i} < T_{D_i} \quad \text{(pre-exposure)} \\
1, & \text{if } T_{A_i} \geq T_{D_i} \quad \text{(post-exposure)}.
\end{cases}
\]
This indicator is used to assign each AE to either the pre- or post-exposure window for drug \( i \). The window is defined symmetrically around \( T_{D_i} \), spanning \( (T_{D_i} - \Delta, T_{D_i} + \Delta) \), and is split into pre-exposure \( (T_{D_i} - \Delta, T_{D_i}) \) and post-exposure \( (T_{D_i}, T_{D_i} + \Delta) \) periods.

\subsection{Assumptions}\label{subsec:assumption}
The statistical validity of our within-subject estimator for the effect of drug \( i \) on the risk of suicide-related events depends on the following assumptions:

\noindent \textbf{1. Temporal Independence.} We assume that the timing of the AE, \( T_{A_i} \), is conditionally independent of the timing of drug exposure, \( T_{D_i} \), given values of the event and exposure indicators:
\[
T_{A_i} \perp T_{D_i} \mid A_i, D_i.
\]
This assumption ensures that observed associations are not confounded by reverse causality. It may be violated for psychiatric drugs, where treatment may be initiated in response to a suicide-related event—resulting in \( T_{A_i} = T_{D_i} \). To address this, we exclude such observations and conduct sensitivity analyses excluding events within one week of exposure. Notably, this bias is unidirectional: it can create spurious protective associations but cannot inflate risk associations.

\noindent \textbf{2. Perfect Compliance.} We assume that patients who fill a prescription take the drug as prescribed. Since prescription fill data reflect dispensing rather than ingestion, this assumption is untestable, but standard in pharmacoepidemiologic analyses. We note however, that it is a more stringent criterion relative to using prescriptions, which may or may not be filled by patient.

\noindent \textbf{3. Exposure Anchoring and Event Uniqueness.} For each drug–subject pair, we consider only one AE occurring in the symmetric window around the first drug fill. This approach avoids complications from repeated events and ensures clear, well-defined exposure and outcome timing.

\noindent \textbf{4. Model Specification.} The probability of a suicide-related event is modeled using a logistic regression:
\begin{equation}\label{eq:logit}
logit \left[ \Pr(A_i = 1 \mid x_i, \mathbf{X}) \right] = \alpha + x_i \theta_i + \mathbf{X}^\top \boldsymbol{\beta}, 
\end{equation}
where \( \theta_i \) is the effect for drug \( i \), \( \mathbf{X} \) is a vector of covariates assumed constant across windows, and \( \boldsymbol{\beta} \) is the corresponding coefficient vector. The intercept \( \alpha \) represents baseline risk. The adjusted odds ratio simplifies algebraically to:
\begin{equation}\label{eq:adj_odds}
\text{OR}_{\text{adj}, i} = \frac{\Pr(A_i = 1 \mid x_i = 1, \mathbf{X})}{\Pr(A_i = 1 \mid x_i = 0, \mathbf{X})} = \frac{\exp(\alpha + \theta_i + \mathbf{X}^\top \boldsymbol{\beta})}{\exp(\alpha + \mathbf{X}^\top \boldsymbol{\beta})} = \exp(\theta_i).    
\end{equation}

Thus, \( \exp(\theta_i) \) reflects the within-subject, covariate-adjusted odds ratio for the effect of drug \( i \) on suicide risk. An odds ratio greater than one ($\text{OR}_{\text{adj}, i}> 1$) indicates increased risk, while $\text{OR}_{\text{adj}, i}<1$ suggests a protective association. These are the targets of inference in our framework.

\subsection{Bayesian Hierarchical Model}

We now present a fully Bayesian hierarchical model for estimating drug-specific associations with suicide-related outcomes. This model builds upon the logistic mixed-effects framework described in \cite{Gibbons2019}, but formalizes it within a rigorous Bayesian paradigm. It also expands the original empirical Bayes structure by incorporating sparsity-inducing priors and a co-prescription-informed covariance model to account for similarities among drugs.

This model refines and extends the conditional logit structure introduced in Assumption 5 of the previous section. As previously discussed, the data are stratified by covariate combinations. For each stratum \( j \) of drug \( i \), let \( \mathbf{Z}_{ij} \in \mathbb{R}^{p} \) denote the vector of covariates other than exposure (age, sex and age-sex interaction), and let \( x_{ij} \in \{0,1\} \) be the main exposure indicator, indicating whether the suicide-related event occurred in the post-exposure window. We define the full covariate vector as \( \mathbf{x}_{ij} = [\mathbf{Z}_{ij}^\top,\, x_{ij}]^\top \in \mathbb{R}^{p+1} \). Let \( Y_{ij} \) be the number of suicide-related events in stratum \( j \), and \( m_{ij} \) the number of claimants. For each drug \( i \in \{1, \ldots, N\} \), the observed outcome is modeled as:
\begin{equation}
Y_{ij} \sim \text{Binomial}(m_{ij}, p_{ij}),
\end{equation}
with the logit-transformed probability specified as:
\begin{equation}
\text{logit}(p_{ij}) = \mathbf{x}_{ij}^\top \boldsymbol{\beta} + \mathbf{x}_{ij}^\top \boldsymbol{\theta}_i.
\end{equation}

Here, \( \boldsymbol{\beta} \in \mathbb{R}^{p+1} \) is a vector of fixed effects shared across all drugs, capturing population-level effects of covariates. The vector \( \boldsymbol{\theta}_i \in \mathbb{R}^{p+1} \) represents the drug-specific random effects, allowing associations with the outcome to vary across drugs. Each covariate, including the post-exposure indicator, contributes both a fixed and a drug-specific random component to the model.

Our primary inferential target is the final component of \( \boldsymbol{\theta}_i \), denoted \( \theta_{i,x} \), which corresponds to \(x_{ij}\). This parameter represents the adjusted, within-subject log-odds ratio for suicide risk associated with drug \( i \), accounting for covariate stratification and baseline heterogeneity (see \ref{eq:adj_odds} in section \ref{subsec:assumption}).

To define a structured prior for the drug-specific effects, we introduce a latent vector \( \boldsymbol{\gamma}_i = (\gamma_{i,1}, \ldots, \gamma_{i,p}, \gamma_{i,x})^\top \in \mathbb{R}^{p+1} \) for each drug such that:
\begin{equation}
\boldsymbol{\theta}_i = (\gamma_{i,1}, \ldots, \gamma_{i,p}, \tilde{\gamma}_{i,x})^\top.
\end{equation}
The first \( p \) elements of \( \boldsymbol{\theta}_i \) are set equal to the corresponding components of \( \boldsymbol{\gamma}_i \), while the final element is modified using a spike-and-slab prior. Specifically, we define:
\begin{align}
\tilde{\gamma}_{i,x} &= \delta_i \cdot \gamma_{i,x}, \label{eq:slab_indicator} \\
\delta_i &\sim \text{Bernoulli}(\pi), \\
\pi &\sim \Pi,
\end{align}
where \( \delta_i \in \{0,1\} \) is a binary indicator denoting whether the exposure effect \( \theta_{i,x} \) is nonzero, and \( \Pi \) is a prior distribution on the inclusion probability \( \pi \), such as a Beta prior. This spike-and-slab prior imposes sparsity on the random slopes for the exposure variable and enables variable selection at the drug level.

To capture correlation across drugs, we model the collection of latent vectors \( \boldsymbol{\gamma}_i \) using a matrix normal prior. Let \( \boldsymbol{\gamma} \in \mathbb{R}^{N(p+1)} \) denote the stacked vector:
\(
\boldsymbol{\gamma} = (\boldsymbol{\gamma}_1^\top \boldsymbol{\gamma}_2^\top \hdots \boldsymbol{\gamma}_N^\top)^\top,
\)
and assume:
\begin{equation}
\boldsymbol{\gamma} \sim \mathcal{N}\left(\mathbf{0}, \Sigma_D \otimes \Sigma_\gamma\right),
\end{equation}
where \( \Sigma_D \in \mathbb{R}^{N \times N} \) is a covariance matrix that captures dependencies among drugs, and \( \Sigma_\gamma \in \mathbb{R}^{(p+1) \times (p+1)} \) is the within-drug covariance matrix of the random effects. The matrix \( \Sigma_D \) is known and its structure is informed by co-prescription frequencies as described in the next subsection.

To ensure flexibility in the prior for \( \Sigma_\gamma \), we place a log-Cholesky prior over its Cholesky factor. Specifically, if \( L \) is the lower-triangular Cholesky decomposition of \( \Sigma_\gamma \), we assume that the logarithms of its diagonal elements and the unconstrained elements below the diagonal are independently distributed, following the log-Cholesky specification introduced by \cite{Barnard2000} and further developed by \cite{Danaher2011}. This formulation avoids the need for positive-definiteness constraints during sampling and provides a parameterization that is both computationally efficient and interpretable.

\subsection{Co-Prescription-Informed Covariance Matrix \( \Sigma_D \)}

In our hierarchical model, the drug-level covariance matrix \( \Sigma_D \) captures structural relationships between drugs based on their real-world co-prescription behavior. We construct \( \Sigma_D \) using co-prescription frequencies extracted from administrative claims data. We have used three distinct methods to derive the covariance matrix \( \Sigma_D \): conditional probability similarity, Pearson correlation, and tetrachoric correlation. Each of these methods defines pairwise association between drugs \( A \) and \( B \) from different statistical perspectives, and are computable from observed data.

\noindent \textbf{Method 1: Conditional Probability-Based Similarity.}
Let \( A \) and \( B \) denote two drugs. Define \( P(A) \) as the probability that a patient is prescribed Drug A, and \( P(A \cap B) \) as the probability that a patient is prescribed both drugs. The conditional probability that Drug B is co-prescribed given Drug A is:
\(
P(B \mid A) = \frac{P(A \cap B)}{P(A)}.
\)
To ensure symmetry, we average the forward and reverse conditional probabilities to define a symmetric similarity score:
\[
\rho_{AB}^{(1)} = \frac{1}{2} \left( \frac{P(A \cap B)}{P(A)} + \frac{P(A \cap B)}{P(B)} \right).
\]

In terms of observed co-prescription counts:
\[
\rho_{AB}^{(1)} = \frac{1}{2} \left( \frac{n_{AB}}{n_A} + \frac{n_{AB}}{n_B} \right),
\]
where \( n_{AB} \) is the number of patients prescribed both drugs, and \( n_A \), \( n_B \) are the numbers of patients prescribed Drug A and Drug B, respectively.

\noindent \textbf{Method 2: Pearson Correlation for Binary Co-Prescription Indicators.}

Define binary indicators \( I_A \) and \( I_B \), where \( I_A = 1 \) if a patient was prescribed Drug A and \( 0 \) otherwise, and similarly for \( I_B \). The Pearson correlation between these indicators is:
\[
\rho_{AB}^{(2)} = \frac{P(A \cap B) - P(A)P(B)}{\sqrt{P(A)(1 - P(A)) \cdot P(B)(1 - P(B))}}.
\]

This measures linear association between co-prescriptions. Using frequency data:
\[
\rho_{AB}^{(2)} = \frac{n_{AB}/n - (n_A/n)(n_B/n)}{\sqrt{(n_A/n)(1 - n_A/n)(n_B/n)(1 - n_B/n)}},
\]
where \( n \) is the total number of patients.

\noindent \textbf{Method 3: Tetrachoric Correlation.}
The tetrachoric correlation estimates the latent correlation between two unobserved, normally distributed variables that are dichotomized to generate the observed binary prescription indicators. Specifically, it assumes:
\[
I_A = \mathbb{I}(Z_A > \tau_A), \quad I_B = \mathbb{I}(Z_B > \tau_B),
\]
where \( Z_A \) and \( Z_B \) are jointly normal with zero means, unit variances, and correlation \( \rho_{AB}^{(3)} \).

This approach differs fundamentally from methods based on observed binary variables, such as conditional probability or Pearson correlation, by modeling co-prescription as a thresholded manifestation of underlying continuous propensities. Because correlations between continuous latent variables are generally stronger than those between their dichotomized counterparts, tetrachoric correlation can reveal deeper patterns of association. It is well-suited to capturing both complementarity and substitution in prescribing behavior, even in settings where the marginal prescription frequencies are highly imbalanced or sparse.

Let \( a \), \( b \), \( c \), and \( d \) be the observed cell counts in the 2×2 co-prescription table:

\[
\begin{array}{c|cc}
    & \text{Drug B Prescribed} & \text{Drug B Not Prescribed} \\
    \hline
    \text{Drug A Prescribed}     & a & b \\
    \text{Drug A Not Prescribed} & c & d \\
\end{array}
\]
Then an approximation to the tetrachoric correlation is given by \cite{Sridharan1952}:
\[
\rho_{AB}^{(3)} \approx \cos\left( \frac{\pi}{1 + \sqrt{(a \cdot d)/(b \cdot c)}} \right),
\]
as discussed in educational literature on psychometric correlation coefficients. This formula provides a fast and stable approximation under a bivariate probit model.


\noindent \textbf{Positive Definiteness.} Although the entries of the co-prescription matrix \( \Sigma_D \) are defined via pairwise association measures, this construction does not, in general, guarantee that the resulting matrix is positive definite. In practice, however, our Bayesian hierarchical model requires \( \Sigma_D \) to be a valid covariance (or correlation) matrix. We therefore replace the raw co-prescription matrix by its nearest symmetric positive definite approximation, obtained by an eigenvalue adjustment that sets nonpositive eigenvalues to a small positive constant and rescales if necessary. This type of ``nearest positive definite" correction is standard in multivariate analysis and numerical linear algebra (see \cite{Higham2002}).

\begin{figure}
    \centering
    \includegraphics[width=1.1\linewidth]{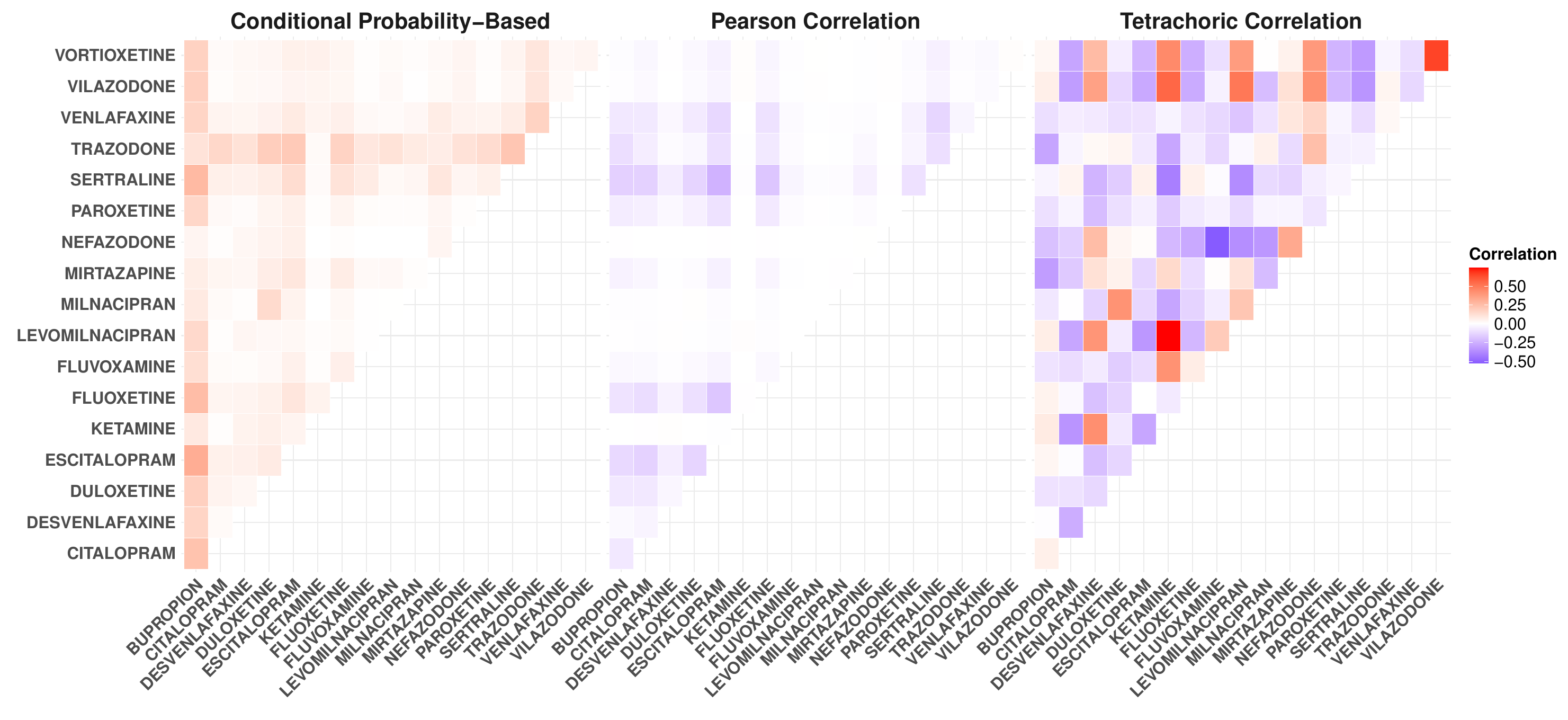}
    \caption{\footnotesize Co-prescription similarity matrices derived using three methods: (Left) conditional probability-based similarity, (Center) Pearson correlation of binary indicators, and (Right) tetrachoric correlation.}
    \label{fig:coprescription_matrix}
\end{figure}

\noindent \textbf{Interpretation and Implications.}
Figure \ref{fig:coprescription_matrix} displays the co-prescription matrices estimated by each of the three methods. High positive entries in the conditional probability or Pearson correlation matrices indicate that a pair of drugs is often prescribed together and may reflect therapeutic complementarity or clinical co-use. Negative values, particularly under Pearson or tetrachoric correlation, suggest that drugs may substitute for one another or belong to mutually exclusive treatment pathways.

The tetrachoric correlation matrix stands out for producing a broader range of values and stronger magnitudes of associations. Specifically, it exhibits a smaller number of strong positive correlations and a larger number of moderate to strong negative correlations compared to the other two methods. This pattern reflects its ability to capture latent substitution patterns that may be obscured in frequency-based methods. 

When used as the covariance matrix \( \Sigma_D \) in our Bayesian hierarchical model, these differences can significantly influence inference. A \( \Sigma_D \) dominated by strong positive correlations encourages information sharing among drugs and may increase power to detect weak but consistent signals. Conversely, a matrix with many negative entries, as in the tetrachoric case, imposes partial decorrelation across drug effects. This may sharpen selection by preventing the borrowing of spurious signal across pharmacologically distinct medications, but it can also reduce borrowing and thus reduce sensitivity in sparsely observed drugs.

\noindent \textbf{Choice of Correlation Metric Affects Both the Estimated Strength and the Direction of Drug-Drug Associations.}
Upon examining the resulting co-prescription matrices (Figure~\ref{fig:coprescription_matrix}), we observed that both the sign and the magnitude of drug-drug associations can change depending on the correlation measure used. For example, the Pearson correlation between two drugs may be positive, while the corresponding tetrachoric correlation may be negative. This phenomenon is not surprising and arises due to fundamental differences in what each correlation metric captures. The Pearson correlation for binary variables measures linear dependence between the observed 0--1 indicators, which is heavily influenced by marginal prevalence. If one drug is very rarely prescribed compared to another, the Pearson correlation can be artificially small or even positive despite underlying competition or substitution. In contrast, the tetrachoric correlation estimates the latent correlation between unobserved continuous variables that, after thresholding, generate the observed binary outcomes. Tetrachoric correlations attempt to correct for imbalances in marginal distributions and can reveal latent antagonistic relationships even when observed prescription patterns appear weakly positively correlated.  Furthermore, conditional probability-based measures are asymmetric by construction (i.e., \( P(A \text{ prescribed} \mid B \text{ prescribed}) \neq P(B \text{ prescribed} \mid A \text{ prescribed}) \)), leading to yet another perspective on association that emphasizes directionality rather than symmetric dependence. Given these conceptual differences, it is natural that the choice of correlation metric affects the estimated strength and the direction of drug-drug associations.

\subsection{False Discovery Rate Control and Variable Selection}

Our goal is to identify drugs whose exposure is significantly associated with suicide-related outcomes, either harmful or protective, while controlling the false discovery rate (FDR) across many simultaneous hypotheses. We adopt a Bayesian decision-theoretic approach to variable selection that leverages posterior inclusion probabilities and a principled loss minimization criterion.

\textbf{Prior on inclusion probability \(\pi\).}
We assign a Beta prior to the global inclusion probability \( \pi \), which governs the sparsity level across all drugs. Following the recommendation of \citet{ScottBerger2010}, we use the default choice:
\begin{equation}
    \pi \sim \text{Beta}(1, 1).
\end{equation}
This uniform prior reflects prior ignorance about the number of non-null effects and allows the posterior to adapt to the level of sparsity in the data. Unlike hyperpriors such as \( \text{Beta}(1, N) \), which strongly favor sparsity, the Beta(1,1) prior is non-informative and avoids imposing excessive shrinkage in situations where multiple true signals may exist. It also induces automatic multiplicity correction by shrinking posterior inclusion probabilities when evidence is weak or marginal likelihoods are flat.

\textbf{Bayesian FDR and decision-theoretic framework.}
Using the notation in the spike and slab model in (\ref{eq:slab_indicator}), \(\delta_i \in \{0,1\}\) denote the latent indicator of the drug level random effect \(\theta_{i, x}\) of $x_{ij}$ (see section \ref{sec:model} for details) specifying whether drug \(i\) is truly associated with the outcome of interest, where \(\delta_i = 1\) indicates the presence of a true nonzero effect and \(\delta_i = 0\) indicates no effect. Additionally, let \(d_i \in \{0,1\}\) represent the selection decision for drug \(i\), such that drug \(i\) is selected as significantly associated if \(d_i = 1\).

The following quantities define the essential components for the Bayesian decision-theoretic approach. Let \(V\) denote the number of false discoveries, which is the number of drugs incorrectly identified as associated. Let \(R\) represent the total number of drugs selected as significant. Let \(T\) indicate the total number of drugs that truly have nonzero effects. Finally, let \(F\) be the number of false non-discoveries, corresponding to true signals that are not selected. Formally, these quantities are defined as
\begin{equation}
    V = \sum_{i=1}^N d_i(1-\delta_i), \quad
R = \sum_{i=1}^N d_i, \quad
T = \sum_{i=1}^N \delta_i, \quad
F = \sum_{i=1}^N (1 - d_i)\delta_i.
\end{equation}

Following \citet{Muller2007}, we define the false discovery rate (FDR) and the false non-discovery rate (FNR) for a given selection rule \(\mathbf{d} = (d_1,\ldots,d_N)\) as follows:
\[
\text{FDR} = \frac{V}{\max(R,1)}, \quad \text{and} \quad \text{FNR} = \frac{F}{\max(N-R,1)}.
\]

From a Bayesian perspective, it is natural to consider posterior expected values of these rates, leading to the definitions of posterior expected false discovery rate (\(\overline{\text{FDR}}\)) and posterior expected false non-discovery rate (\(\overline{\text{FNR}}\)):
\begin{equation}
\overline{\text{FDR}} = \mathbb{E}(\text{FDR}\mid\text{data}),\quad
\overline{\text{FNR}} = \mathbb{E}(\text{FNR}\mid\text{data}).    
\end{equation}

To control the posterior expected false discovery rate at a predetermined level \(\alpha_R\), the optimal selection rule minimizes the posterior expected false non-discovery rate subject to the constraint that the posterior expected false discovery rate does not exceed \(\alpha_R\). Formally, this optimization problem is expressed as:

\begin{equation}
\mathbf{d}^* = \arg\min_{\mathbf{d}\in\{0,1\}^N}\overline{\text{FNR}}\quad \text{subject to}\quad\overline{\text{FDR}}\leq \alpha_R.
\end{equation}

\citet{Muller2007} showed that the solution to this optimization problem is given by a simple thresholding rule based on posterior inclusion probabilities. Specifically, defining the posterior inclusion probability for drug \(i\) as \(\text{PIP}_i = \Pr(\delta_i = 1 \mid \text{data})\), the optimal decision rule is
\begin{equation}
d_i^* = \mathbb{I}(\text{PIP}_i \geq t_{\alpha_R}),    
\end{equation}
where the optimal threshold \(t_{\alpha_R}\) is computed as
\begin{equation}
t_{\alpha_R} = \arg\min_{t \in [0,1]} \left\{\frac{\sum_{i=1}^N \text{PIP}_i\,\mathbb{I}(\text{PIP}_i<t)}{\max\left(N - \sum_{i=1}^N\mathbb{I}(\text{PIP}_i\geq t),1\right)}: \frac{\sum_{i=1}^N (1-\text{PIP}_i)\mathbb{I}(\text{PIP}_i\geq t)}{\max\left(\sum_{i=1}^N\mathbb{I}(\text{PIP}_i\geq t),1\right)} \leq \alpha_R\right\}.    
\end{equation}

In practice, we solve this optimization numerically by performing a grid search over potential threshold values \(t\), selecting the threshold \(t_{\alpha_R}\) that minimizes the posterior expected false non-discovery rate while satisfying the FDR constraint.

This posterior-based decision rule avoids reliance on \( p \)-values and accommodates model uncertainty and correlation across tests. It unifies variable selection and error control, and naturally identifies both harmful (positive) and protective (negative) drug effects. The selected drugs are those with the strongest posterior evidence of association, and their effects are fully interpretable via their estimated log-odds ratios \( \theta_{i, x} \).



\subsection{Simulation Study}
We conducted simulation studies to evaluate the performance of our proposed fully Bayesian spike-and-slab approach with Bayesian false discovery rate (FDR) control, in comparison to existing empirical Bayes approaches using Bonferroni and Benjamini--Hochberg (BH) corrections. The simulation study serves two main purposes. First, we aim to compare our method against existing approaches when co-prescription information is not available, which represents a typical situation in many observational healthcare datasets. Second, we demonstrate the added value of incorporating co-prescription information through a structured covariance matrix \( \Sigma_D \) when such information is available.

In the first scenario, where co-prescription information is unavailable, we compare our proposed fully Bayesian method with Bayesian FDR control to the empirical Bayes approach described in \cite{Gibbons2019}, using Bonferroni correction for FDR control, and to the same empirical Bayes approach using BH correction. This allows us to evaluate the advantages of the fully Bayesian approach over standard frequentist multiple-testing corrections, even without the co-prescription matrix. We use \(\Sigma_D\) as an identity matrix in this scenario.

In the second scenario, we consider the setting where co-prescription information is available. In this case, we evaluate four different methods: (1) the empirical Bayes approach by \cite{Gibbons2019} with Bonferroni correction, (2) the same empirical Bayes approach with BH correction, (3) the fully Bayesian spike-and-slab method without incorporating co-prescription information (i.e., \( \Sigma_D \) is the identity matrix), and (4) the fully Bayesian spike-and-slab method with co-prescription information (proposed method, using a structured \( \Sigma_D \)). This comparison highlights how ignoring co-prescription relationships in the modeling process, as done in the empirical Bayes methods and in the Bayesian model without \( \Sigma_D \), can lead to suboptimal drug selection and poor FDR control. In contrast, incorporating such information leads to more powerful and accurate identification of important drug associations.

\subsubsection{Simulation Design}

In both simulation settings, the data were generated under the same model specification as used in our data analysis. Specifically, the generative model was a binomial logistic regression, where for each drug \(i\) and individual \(j\), the response variable \(Y_{ij} \sim \text{Binomial}(m_{ij}, p_{ij})\) with log-odds defined as
\begin{equation}
    \text{logit}(p_{ij}) = \beta_1 + \beta_2 Z_{ij,1} + \beta_3 Z_{ij,2} + \beta_4 Z_{ij,1}Z_{ij,2} + \beta_5 x_{ij} + \theta_{i,1} + \theta_{i,x} x_{ij},
\end{equation}
where \(Z_{ij,1}\) is the binary indicator for age greater than 18, \(Z_{ij,2}\) is the binary indicator for female sex, and \(x_{ij}\) is the binary time indicator (after vs. before prescription). The drug-specific random intercept and slope for \(x_{ij}\) are denoted by \(\theta_{i,1}\) and \(\theta_{i,x}\), respectively, with a spike-and-slab prior imposed on \(\theta_{i,x}\) to allow for data-driven variable selection.

In practice, we would typically add random-effects for each of the covariates (e.g., age and sex) so that they would take on drug-specific values; however, for the purpose of the simulation studies computational tractability and our primary focus on the drug effects we include then as fixed-effects only.

\noindent \textbf{Preserving Rare-Event Structure.} We used the posterior means of \(\beta_1, \dots, \beta_5\) and \(\theta_{i,1}\) from the fitted model to the real data as the true values for simulations. The true coefficient values were set as follows: \(\beta_1 = -9.36\), \(\beta_2 = 1.10\), \(\beta_3 = -0.212\), \(\beta_4 = -0.823\), and \(\beta_5 = 0.078\). For \(m_{ij}\), we also used the values from the real data. We used the actual values from the real dataset to preserve the inherent data complexity and rare-event characteristics.

We present results from two distinct simulation scenarios:

\noindent \textbf{Simulation Scenario 1:}
We considered a large-scale scenario involving 922 drugs, with 90 drugs having true effects on suicide risk. The values for \(\theta_{i,x}\) were set to mimic the effect sizes observed in the real data: 10 drugs had \(\theta_{i,x} = -1.00\), 20 drugs had \(\theta_{i,x} = -0.75\), 30 drugs had \(\theta_{i,x} = -0.50\) (protective effects), and 30 drugs had \(\theta_{i,x} = 0.50\) (harmful effects). The remaining 832 drugs had \(\theta_{i,x} = 0\).

\noindent \textbf{Simulation Scenario 2:}We considered a smaller-scale but more challenging scenario involving 100 randomly selected drugs from the full dataset, of which 20 had true effects. Among these, three drugs had very strong protective effects with \(\theta_{i,x} = -0.75\), and 17 drugs had moderate protective effects with \(\theta_{i,x} = -0.50\). To explicitly evaluate the advantage of incorporating co-prescription information, we included a structured co-prescription matrix \(\Sigma_D\) in this scenario. This matrix was symmetric, with diagonal entries equal to 1 and off-diagonal entries uniformly distributed between 0.25 and 0.40 among the first 30 drugs (which included all 20 important drugs). All other off-diagonal entries were set to zero, thereby simulating a realistic yet sparse drug interaction network.

\noindent \textbf{Performance Metrics.} The performance of each method was evaluated based on the number of selected drugs, statistical power (the proportion of truly important drugs selected), and the false discovery rate (the proportion of selected drugs that were false positives). We report the median and median absolute deviation (MAD) of these metrics from multiple simulated datasets.

\subsubsection{Simulation Study Results}

\paragraph{Scenario 1 Results (Target FDR $\leq 0.05$):}
Table~\ref{tab:scenario1_results} summarizes the performance across methods. The empirical Bayes approach with Bonferroni correction achieved excellent FDR control (zero false discoveries), but at the cost of substantially reduced power (median power = 0.65). Replacing Bonferroni with BH improved power (0.81) with a modest increase in FDR (0.02). In contrast, our fully Bayesian spike-and-slab approach offered the best balance, recovering the largest number of true associations (median power = 0.90) while keeping the FDR within the desired threshold (0.04). The reduced variability in both power and FDR (as seen from the MADs) suggests a more stable and efficient variable selection process under our proposed model.

\begin{table}[ht]
\centering
\caption{Simulation Scenario 1 Results (Targeted FDR $\leq 0.05$)}
\renewcommand{\arraystretch}{0.8}
\fontsize{11}{11}\selectfont
\begin{tabular}{lccc}
\hline
Method & Number Selected & Power & FDR \\
& Median (MAD) & Median (MAD) & Median (MAD)\\
\hline
Bonferroni & 58 (2) & 0.65 (0.02) & 0 (0) \\
BH & 74 (2) & 0.81 (0.02) & 0.02 (0.01) \\
Spike-and-Slab (Proposed) & 84 (1) & 0.90 (0.01) & 0.04 (0.01) \\
\hline
\end{tabular}
\label{tab:scenario1_results}
\end{table}

\begin{table}[ht]
\centering
\caption{Simulation Scenario 2 Results (Targeted FDR $\leq 0.05$)}
\renewcommand{\arraystretch}{0.8}
\fontsize{11}{11}\selectfont
\begin{tabular}{lccc}
\hline
Method & Number Selected & Power & FDR \\
& Median (MAD) & Median (MAD) & Median (MAD)\\
\hline
Bonferroni & 8 (1) & 0.35 (0.05) & 0 (0) \\
BH & 14 (2) & 0.60 (0.05) & 0.13 (0.05) \\
Spike-and-Slab & 17 (2) & 0.80 (0.10) & 0.06 (0.06) \\
Spike-and-Slab with Co-prescription & \multirow{2}{*}{19 (1)} & \multirow{2}{*}{0.95 (0.05)} & \multirow{2}{*}{0 (0)} \\
(Proposed) & & & \\
\hline
\end{tabular}
\label{tab:scenario2_results_fdr05}
\end{table}

\begin{table}[ht]
\centering
\caption{Simulation Scenario 2 Results (Targeted FDR $\leq 0.15$)}
\renewcommand{\arraystretch}{0.8}
\fontsize{11}{11}\selectfont
\begin{tabular}{lccc}
\hline
Method & Number Selected & Power & FDR \\
& Median (MAD) & Median (MAD) & Median (MAD)\\
\hline
Bonferroni & 10 (1) & 0.45 (0.05) & 0.10 (0.07) \\
BH & 21 (3) & 0.75 (0.05) & 0.26 (0.06) \\
Spike-and-Slab & 22 (2) & 0.90 (0.10) & 0.17 (0.12) \\
Spike-and-Slab with Co-prescription & \multirow{2}{*}{22 (1)} & \multirow{2}{*}{1.00 (0)} & \multirow{2}{*}{0.09 (0.04)} \\
(Proposed) & & & \\
\hline
\end{tabular}
\label{tab:scenario2_results_fdr15}
\end{table}

\paragraph{Scenario 2 Results:}
In Simulation Scenario 2, we evaluated performance under two different target false discovery rates (FDRs), namely \(\alpha = 0.05\) and \(\alpha = 0.15\). This dual-threshold evaluation was motivated by practical considerations arising in high-dimensional sparse settings, where the number of true signals is small relative to the total number of hypotheses. In our simulation, only 20 drugs were designated as truly associated with suicide risk. When the number of true discoveries is modest, even a single false positive can cause the observed FDR to exceed a stringent threshold such as 0.05. For example, if a method successfully selects 15 true signals but also incurs one false discovery, the empirical FDR would be approximately \(1/16 \approx 0.0625\), exceeding the 0.05 limit. As a result, methods that perform reasonably well in terms of power can still appear to violate strict FDR control purely due to sampling variability or the inherent discreteness of the discovery process. To better capture the practical operating characteristics of the methods, we therefore examined performance both at a stringent FDR level (\(\alpha = 0.05\)) and a more lenient level (\(\alpha = 0.15\)). This approach allows a more comprehensive evaluation of the trade-off between discovery power and false discovery control in sparse, low-signal environments, and mirrors similar practices in empirical studies of high-dimensional screening.

Tables~\ref{tab:scenario2_results_fdr05} and \ref{tab:scenario2_results_fdr15} summarize the results under FDR thresholds of 0.05 and 0.15, respectively. Under the more stringent threshold of 0.05, the Bonferroni method again yielded perfect FDR control, but its low power (0.35) severely limited utility. The BH procedure offered better power (0.60) but came with a relatively high FDR (0.13). Our spike-and-slab method without co-prescription information improved both metrics (power = 0.80, FDR = 0.06), while incorporating the co-prescription matrix further enhanced power (0.95) and simultaneously reduced false discoveries (FDR = 0).

At the relaxed FDR level of 0.15, all methods selected more drugs, but the differences between approaches became even more pronounced. BH's FDR rose to 0.26, while our fully Bayesian method with co-prescription maintained an FDR well within bounds (0.09) and achieved perfect power (1.00). This result demonstrates the robustness of our model under various FDR settings and emphasizes the substantial benefits of incorporating external biological structure through \( \Sigma_D \).

\noindent \textbf{Choice of \(\alpha_R\) in Different Settings:}
In our simulation studies and real data analysis, we carefully calibrated the choice of \(\alpha_R\), the target upper bound for the posterior expected false discovery rate (\(\overline{\text{FDR}}\)). Although \(\alpha_R\) is typically set equal to the desired FDR control level, our empirical observations revealed important nuances depending on the dimensionality and sparsity of the problem.
In Simulation Scenario 1, which involved a very high-dimensional setting with 922 drugs and extreme sparsity (only 90 true signals), we found that setting \(\alpha_R = 0.05\) often led to slight exceedance of the nominal 0.05 frequentist FDR level. Due to the high degree of sparsity, the posterior distributions for the inclusion indicators tended to be conservative, and using \(\alpha_R = 0.05\) did not sufficiently compensate for the discreteness and variability introduced by extremely rare signals. To address this, we employed a more stringent choice of \(\alpha_R = 0.02\), which empirically resulted in better alignment between the targeted and observed FDR levels, maintaining the realized FDR close to 0.05.
For consistency, in the real data analysis—where the dimensionality and sparsity were comparable to Scenario 1—we similarly set \(\alpha_R = 0.02\) to achieve effective control at the nominal 0.05 FDR level.
In contrast, Simulation Scenario 2 involved a more moderate dimensionality (100 drugs) and a larger proportion of true signals (20 associated drugs). In this setting, the posterior calibration was less sensitive to small perturbations, and we observed that setting \(\alpha_R\) equal to the desired FDR target (either 0.05 or 0.15) adequately controlled the FDR without the need for further adjustment.

These simulation studies clearly demonstrate that our proposed Bayesian spike-and-slab method, particularly when leveraging co-prescription information, delivers superior accuracy, interpretability, and control over false discoveries in high-dimensional drug screening settings. The method remains robust across different effect sizes, sparsity levels, and FDR targets, thereby offering a compelling tool for observational pharmacovigilance applications.

\section{Application to Real Data}

In this section, we present the findings from applying the proposed Bayesian spike-and-slab model with FDR conamatrol to the real-world healthcare claims data. The results are organized into two subsections based on the availability of co-prescription information and the specific analysis goals. First, we report the analysis results for the full dataset consisting of 922 drugs. This is the same example used in \cite{Gibbons2019}. In this setting, co-prescription information was not incorporated into the model, as such data were not uniformly available across all drugs. We focus on identifying drugs associated with suicide risk or protection using a prior that assumes independence across drug effects. Second, we conduct a focused analysis on a subset of 18 antidepressant drugs for which co-prescription data were available. In this part, we systematically compare drug selection results obtained under three different constructions of the co-prescription-informed covariance matrix: conditional probability-based, Pearson correlation-based, and tetrachoric correlation-based matrices. We assess how incorporating co-prescription structure influences drug selection relative to the model assuming independence and highlight cases where accounting for drug–drug dependencies reveals new signals or changes interpretation.

\subsection{Full Dataset Analysis Without Co-prescription Information}

We first applied the proposed Bayesian spike-and-slab model to the full dataset consisting of \(N = 922\) drugs. Co-prescription information was not incorporated at this stage, and independence across drugs was assumed by setting the co-prescription matrix \(\Sigma_D\) to the identity matrix. Drug selection was based on posterior inclusion probabilities (PIPs), with false discovery rate (FDR) control achieved by setting \(\alpha_R = 0.02\), calibrated to maintain an empirical FDR near \(0.05\).

\begin{figure}[htbp]
\centering
\includegraphics[width=0.7\textwidth]{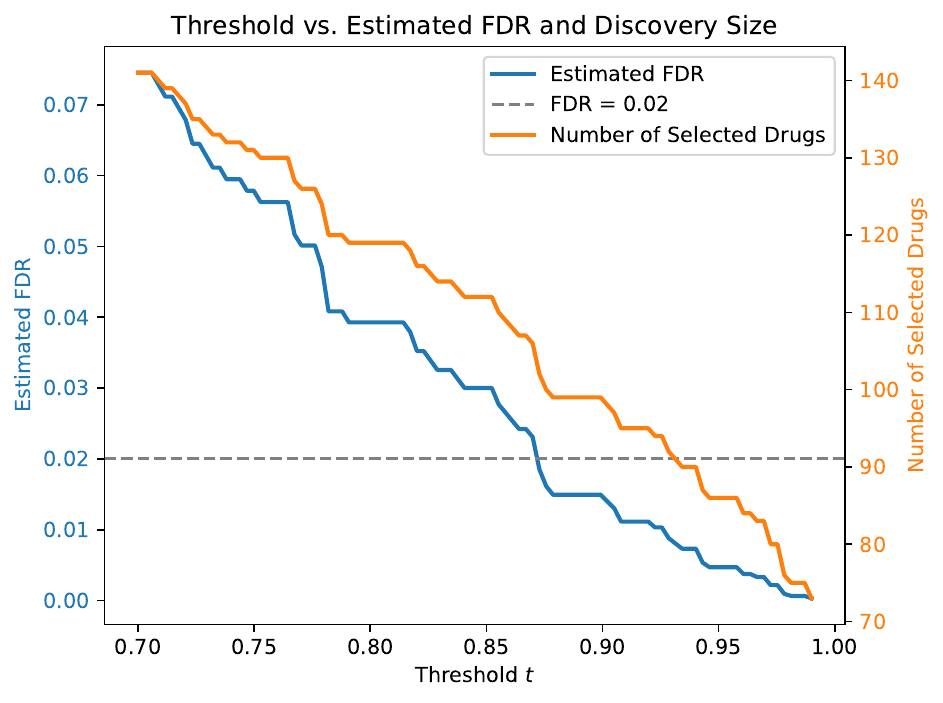}
\caption{\footnotesize Estimated false discovery rate ($\widehat{\mathrm{FDR}}$) and number of selected drugs versus threshold $t$ in the Bayesian spike-and-slab selection. The dashed line marks the targeted $\widehat{\mathrm{FDR}} = 0.02$ level.}
\label{fig:fdr_vs_selection}
\end{figure}

The proposed method identified 92 drugs significantly associated with suicide-related outcomes. (21 were associated with an increase in suicide attempt and 71 were associated with reduced risk).  By comparison \cite{Gibbons2019} found 10 drugs with increased risk and 44 with decreased risk. These results were obtained directly from the fully Bayesian variable selection model without refitting drug-specific models post-selection. Each estimated effect represents the adjusted odds ratio for post-exposure suicide events, adjusting for age, sex, and the age-sex interaction as covariates. Figure~\ref{fig:fdr_vs_selection} illustrates the relationship between the selection threshold $t$, the estimated false discovery rate, and the number of selected drugs, highlighting the trade-off between discovery size and error control. 

To better understand the directionality of associations, we categorized the selected drugs based on the posterior mean of their adjusted odds ratios: drugs with posterior mean greater than 1 were interpreted as associated with an \textit{increased} risk, while drugs with posterior mean less than 1 were interpreted as associated with a \textit{decreased} risk of suicidal events. The full lists are presented in Table~\ref{tab:increased_risk} for drugs associated with increased risk and Table~\ref{tab:decreased_risk} for those associated with decreased risk.  Overall, the identified drug classes were similar to those found by \cite{Gibbons2019}.  Antidepressants, anti-manic and antipsychotic medications were generally associated with decreased risk.  Curiously, these drugs all have black box warnings from the FDA warning of increased suicide risk, a warning that has been hotly debated for the past 20 years (see \cite{Gibbons2007Early, Libby2009Persisting, Busch2014FDA}), who all showed that following the black box warning there was a large reduction in treatment for depressed children and following the black box warning the rate of suicide in children and adolescents increased rather than decreased.   While many of the same drugs previously identified to be associated with increased suicide attempt risk by \cite{Gibbons2019} were also found here, the largest signal for podofilox, an antimitotic drug used to treat genital warts, was not previously identified by \cite{Gibbons2019}.  This increased risk could be due to the social discomfort of having this condition or possibly due to the drug's mechanism which inhibits cell growth and division.  Further studies ae required.  

\begin{table}[ht]
\centering
\caption{Drugs Associated with Increased Risk of Suicide with Posterior Mean ($>1$) and 95\% Credible intervals (CrI). The Adjusted odds ratios are adjusted for age, sex, and age-sex interaction, derived from the Bayesian spike-and-slab variable selection model.}
\label{tab:increased_risk}
\renewcommand{\arraystretch}{0.7}
\fontsize{10}{10}\selectfont
\begin{tabular}{lccc}
\toprule
Drug Name & Posterior Mean & 95\% Lower CrI & 95\% Upper CrI \\
\midrule
Podofilox & 2.009 & 0.935 & 3.467 \\
Fluocinolone Acetonide & 1.783 & 0.995 & 2.708 \\
Acetaminophen/Butalbital/Caffeine & 1.756 & 1.446 & 2.093 \\
Isotretinoin & 1.742 & 1.000 & 2.625 \\
Carisoprodol & 1.691 & 1.349 & 2.003 \\
Alprazolam & 1.664 & 1.523 & 1.787 \\
Hydromorphone Hydrochloride & 1.629 & 1.242 & 2.006 \\
Ethinyl Estradiol/Norgestrel & 1.598 & 1.000 & 2.295 \\
Rizatriptan Benzoate & 1.507 & 1.082 & 1.953 \\
Codeine Phosphate/Promethazine Hydrochloride & 1.485 & 1.164 & 1.742 \\
Clarithromycin & 1.470 & 1.169 & 1.800 \\
Chlorpheniramine Polistirex/Hydrocodone Polistirex & 1.434 & 1.117 & 1.832 \\
Varenicline & 1.418 & 1.000 & 1.793 \\
Montelukast Sodium & 1.335 & 1.000 & 1.599 \\
Prednisone & 1.270 & 1.156 & 1.415 \\
Cyclobenzaprine Hydrochloride & 1.243 & 1.134 & 1.353 \\
Promethazine Hydrochloride & 1.242 & 1.100 & 1.375 \\
Diazepam & 1.242 & 1.092 & 1.357 \\
Acetaminophen/Hydrocodone Bitartrate & 1.241 & 1.167 & 1.310 \\
Amitriptyline Hydrochloride & 1.200 & 1.000 & 1.369 \\
Azithromycin & 1.173 & 1.091 & 1.260 \\
\bottomrule
\end{tabular}
\end{table}

\begin{table}[!ht]
\centering
\caption{\small Drugs Associated with Decreased Risk of Suicide with Posterior Mean ($<1$) and 95\% Credible intervals (CrI). The Adjusted odds ratios are adjusted for age, sex, and age-sex interaction, derived from the Bayesian spike-and-slab variable selection model.}
\label{tab:decreased_risk}
\renewcommand{\arraystretch}{0.45}
\fontsize{8}{8.5}\selectfont
\begin{tabular}{lccc}
\toprule
Drug Name & Posterior Mean & 95\% CrI Lower & 95\% CrI Upper \\
\midrule
Folic Acid & 0.353 & 0.274 & 0.414 \\
Mirtazapine & 0.359 & 0.325 & 0.389 \\
Trazodone Hydrochloride & 0.366 & 0.348 & 0.387 \\
Hydroxyzine Pamoate & 0.366 & 0.338 & 0.401 \\
Disulfiram & 0.383 & 0.288 & 0.487 \\
Naltrexone & 0.403 & 0.331 & 0.460 \\
Aripiprazole & 0.408 & 0.381 & 0.440 \\
Lithium Carbonate & 0.414 & 0.375 & 0.455 \\
Divalproex Sodium & 0.423 & 0.387 & 0.465 \\
Prazosin Hydrochloride & 0.437 & 0.348 & 0.515 \\
Quetiapine Fumarate & 0.445 & 0.416 & 0.479 \\
Carbamazepine & 0.446 & 0.370 & 0.523 \\
Buspirone Hydrochloride & 0.467 & 0.422 & 0.504 \\
Risperidone & 0.471 & 0.434 & 0.510 \\
Asenapine & 0.482 & 0.339 & 0.613 \\
Olanzapine & 0.487 & 0.433 & 0.541 \\
Hydroxyzine Hydrochloride & 0.493 & 0.437 & 0.544 \\
Benztropine Mesylate & 0.493 & 0.418 & 0.562 \\
Oxcarbazepine & 0.498 & 0.441 & 0.565 \\
Clozapine & 0.500 & 0.271 & 0.708 \\
Citalopram Hydrobromide & 0.511 & 0.479 & 0.547 \\
Acamprosate Calcium & 0.514 & 0.416 & 0.621 \\
Lamotrigine & 0.516 & 0.473 & 0.556 \\
Bupropion Hydrochloride & 0.526 & 0.491 & 0.561 \\
Lurasidone Hydrochloride & 0.526 & 0.419 & 0.635 \\
Amlodipine Besylate & 0.545 & 0.445 & 0.635 \\
Paliperidone & 0.553 & 0.360 & 0.737 \\
Guanfacine Hydrochloride & 0.558 & 0.440 & 0.694 \\
Buprenorphine/Naloxone & 0.558 & 0.460 & 0.681 \\
Diphenhydramine Hydrochloride & 0.574 & 0.314 & 0.900 \\
Fluoxetine Hydrochloride/Olanzapine & 0.579 & 0.371 & 0.873 \\
Escitalopram Oxalate & 0.581 & 0.546 & 0.612 \\
Sertraline Hydrochloride & 0.583 & 0.549 & 0.619 \\
Hydralazine Hydrochloride & 0.589 & 0.339 & 1.000 \\
Buprenorphine & 0.602 & 0.399 & 0.819 \\
Glucose Meter & 0.603 & 0.390 & 1.000 \\
Gabapentin & 0.607 & 0.557 & 0.658 \\
Duloxetine Hydrochloride & 0.608 & 0.559 & 0.670 \\
Venlafaxine Hydrochloride & 0.613 & 0.567 & 0.657 \\
Haloperidol & 0.613 & 0.483 & 0.751 \\
Famotidine & 0.618 & 0.482 & 0.746 \\
Pantoprazole Sodium & 0.624 & 0.559 & 0.702 \\
Propranolol Hydrochloride & 0.625 & 0.538 & 0.712 \\
Perphenazine & 0.628 & 0.410 & 0.863 \\
Valsartan & 0.630 & 0.417 & 1.000 \\
Fluoxetine Hydrochloride & 0.644 & 0.607 & 0.681 \\
Lisinopril & 0.650 & 0.567 & 0.749 \\
Levomefolate Calcium & 0.650 & 0.450 & 1.000 \\
Ziprasidone Hydrochloride & 0.661 & 0.585 & 0.741 \\
Doxepin Hydrochloride & 0.663 & 0.553 & 0.784 \\
Sulfacetamide Sodium & 0.666 & 0.483 & 1.000 \\
Ropinirole Hydrochloride & 0.671 & 0.447 & 0.902 \\
Valproic Acid & 0.671 & 0.422 & 1.000 \\
Clonidine Hydrochloride & 0.671 & 0.600 & 0.765 \\
Chlordiazepoxide Hydrochloride & 0.680 & 0.537 & 0.868 \\
Desvenlafaxine & 0.685 & 0.571 & 0.799 \\
Chlorpromazine Hydrochloride & 0.728 & 0.557 & 1.000 \\
Atomoxetine Hydrochloride & 0.738 & 0.576 & 0.897 \\
Polyethylene Glycol 3350 & 0.744 & 0.595 & 1.000 \\
Hydrochlorothiazide & 0.745 & 0.621 & 0.850 \\
Potassium Chloride & 0.749 & 0.616 & 0.891 \\
Metformin Hydrochloride & 0.758 & 0.596 & 0.942 \\
Topiramate & 0.759 & 0.668 & 0.854 \\
Temazepam & 0.812 & 0.725 & 0.899 \\
Cephalexin & 0.815 & 0.754 & 0.885 \\
Metoprolol & 0.823 & 0.730 & 1.000 \\
Methylphenidate & 0.840 & 0.732 & 1.000 \\
Sulfamethoxazole/Trimethoprim & 0.860 & 0.783 & 0.933 \\
Zolpidem Tartrate & 0.862 & 0.801 & 0.924 \\
Lorazepam & 0.864 & 0.805 & 0.924 \\
Amoxicillin/Clavulanate Potassium & 0.898 & 0.818 & 1.000 \\
\bottomrule
\end{tabular}
\end{table}

\begin{table}[!ht]
\centering
\caption{Extra Drugs Detected Compared to \cite{Gibbons2019} with Posterior Mean and 95\% Credible intervals (CrI).}
\label{tab:extra_drugs}
\renewcommand{\arraystretch}{0.63}
\fontsize{9.5}{9.5}\selectfont
\begin{tabular}{lccc}
\toprule
Drug Name & Posterior Mean & 95\% CrI Lower & 95\% CrI Upper \\
\midrule
Amitriptyline Hydrochloride & 1.200 & 1.000 & 1.369 \\
Amoxicillin/Clavulanate Potassium & 0.898 & 0.818 & 1.000 \\
Atomoxetine Hydrochloride & 0.738 & 0.576 & 0.897 \\
Buprenorphine & 0.602 & 0.399 & 0.819 \\
Cephalexin & 0.815 & 0.754 & 0.885 \\
Chlordiazepoxide Hydrochloride & 0.680 & 0.537 & 0.868 \\
Chlorpheniramine Polistirex/Hydrocodone Polistirex & 1.434 & 1.117 & 1.832 \\
Chlorpromazine Hydrochloride & 0.728 & 0.557 & 1.000 \\
Clarithromycin & 1.470 & 1.169 & 1.800 \\
Codeine Phosphate/Promethazine Hydrochloride & 1.485 & 1.164 & 1.742 \\
Diphenhydramine Hydrochloride & 0.574 & 0.314 & 0.900 \\
Ethinyl Estradiol/Norgestrel & 1.598 & 1.000 & 2.295 \\
Famotidine & 0.618 & 0.482 & 0.746 \\
Fluocinolone Acetonide & 1.783 & 0.995 & 2.708 \\
Fluoxetine Hydrochloride/Olanzapine & 0.579 & 0.371 & 0.873 \\
Glucose Meter & 0.603 & 0.390 & 1.000 \\
Hydralazine Hydrochloride & 0.589 & 0.339 & 1.000 \\
Hydrochlorothiazide & 0.745 & 0.621 & 0.850 \\
Hydromorphone Hydrochloride & 1.629 & 1.242 & 2.006 \\
Isotretinoin & 1.742 & 1.000 & 2.625 \\
Levomefolate Calcium & 0.650 & 0.450 & 1.000 \\
Lorazepam & 0.864 & 0.805 & 0.924 \\
Metformin Hydrochloride & 0.758 & 0.596 & 0.942 \\
Methylphenidate & 0.840 & 0.732 & 1.000 \\
Metoprolol & 0.823 & 0.730 & 1.000 \\
Montelukast Sodium & 1.335 & 1.000 & 1.599 \\
Podofilox & 2.009 & 0.935 & 3.467 \\
Polyethylene Glycol 3350 & 0.744 & 0.595 & 1.000 \\
Potassium Chloride & 0.749 & 0.616 & 0.891 \\
Rizatriptan Benzoate & 1.507 & 1.082 & 1.953 \\
Ropinirole Hydrochloride & 0.671 & 0.447 & 0.902 \\
Sulfacetamide Sodium & 0.666 & 0.483 & 1.000 \\
Sulfamethoxazole/Trimethoprim & 0.860 & 0.783 & 0.933 \\
Temazepam & 0.812 & 0.725 & 0.899 \\
Topiramate & 0.759 & 0.668 & 0.854 \\
Valproic Acid & 0.671 & 0.422 & 1.000 \\
Valsartan & 0.630 & 0.417 & 1.000 \\
Varenicline & 1.418 & 1.000 & 1.793 \\
Zolpidem Tartrate & 0.862 & 0.801 & 0.924 \\
\bottomrule
\end{tabular}
\end{table}

\textbf{Additional Signals Detected Beyond \cite{Gibbons2019}.}
Importantly, our method identified a set of additional drugs that were not detected in the empirical Bayes analysis reported by \citet{Gibbons2019}. These extra discoveries are listed in Table~\ref{tab:extra_drugs}. 
The newly detected drugs highlight the enhanced sensitivity of the spike-and-slab framework, which, by fully propagating posterior uncertainty and adaptively shrinking drug-specific effects, is able to uncover associations that may be subtle but clinically meaningful. For example the antidepressant fluoxetine was found to be associated with decreased suicide attempt risk as were the insomnia medications temazepam and zolpidem. The antiepileptic/antimanic drugs topiramate and valproic acid, and the antipsychotic chlorpromazine were also associated with reduced suicide attempt risk.  Conversely, our analysis revealed associations for increased risk for the opioid-based cold medicine chlorpheniramine Polistirex/Hydrocodone Polistirex, the oral contraceptive ethinyl estradiol/norgestrel, the corticosteroid fluocinolone acetonide, and Rizatriptan Benzoate a migrane medication.  While some of these findings may represent dynamic confounding by indication, the majority are either biologically plausible (antidepressant, antimanic, anti psychotic medications and reduced suicide risk) or are prescribed for conditions unrelated to suicide risk (cold medicines, contraceptives and corticosteroids), illustrating the practical value of improved variable selection in pharmacovigilance research.
These findings demonstrate the robustness and sensitivity of our fully Bayesian methodology for high-dimensional drug safety screening, even when no co-prescription structure is incorporated.

\subsection{Antidepressant Drug Analysis and the Impact of Co-prescription Information}

To gain deeper insight into how explicit incorporation of co-prescription information affects drug selection in pharmacovigilance studies, we conducted a focused analysis on a subset of 18 antidepressant medications using our Bayesian spike-and-slab model. We considered three distinct co-prescription matrices—tetrachoric correlation-based, Pearson correlation-based, and conditional probability-based—and compared these results against an independence scenario (no co-prescription information). Given the smaller number of drugs analyzed, we controlled the posterior expected false discovery rate at a more relaxed threshold (\(\alpha_R = 0.15\)) to facilitate practical discovery and interpretation.

Table~\ref{tab:antidepressant_comprehensive} provides a comprehensive summary of adjusted odds ratios with their 95\% credible intervals for each antidepressant under the four considered scenarios. These adjusted odds ratios, derived from the Bayesian spike-and-slab variable selection model, incorporate adjustments for age, sex, and the age-sex interaction, thus accounting for important demographic covariates known to influence suicidal event risks.

\begin{table}[ht]
\centering
\caption{Comprehensive Comparison of Adjusted Odds Ratios for Antidepressant Drugs Under Different Co-prescription Correlation Structures. Adjusted odds ratios (posterior means and 95\% credible intervals) derived from the Bayesian spike-and-slab model, adjusted for age, sex, and age-sex interaction.}
\label{tab:antidepressant_comprehensive}
\renewcommand{\arraystretch}{1}
\fontsize{8.5}{9}\selectfont
\begin{tabular}{lcccc}
\toprule
Drug Name & Tetrachoric & Pearson & Conditional Probability & Independent \\
& \multicolumn{4}{c}{posterior means (95\% credible intervals) of Adjusted Odds Ratios}\\
\midrule
Bupropion Hydrochloride & 0.880 (0.805, 0.954) & 0.863 (0.790, 0.944) & 0.865 (0.796, 0.942) & 0.884 (0.820, 1.000) \\
Citalopram Hydrobromide & 0.849 (0.780, 0.910) & 0.835 (0.767, 0.906) & 0.838 (0.770, 0.909) & 0.856 (0.793, 0.932) \\
Fluoxetine Hydrochloride & 1.047 (1.000, 1.136) & Not selected & Not selected & 1.057 (1.000, 1.152) \\
Fluvoxamine Maleate & 1.313 (1.000, 1.702) & 1.170 (0.996, 1.582) & 1.190 (0.993, 1.611) & 1.253 (1.000, 1.639) \\
Mirtazapine & 0.593 (0.539, 0.658) & 0.578 (0.517, 0.645) & 0.581 (0.530, 0.645) & 0.593 (0.527, 0.658) \\
Paroxetine Hydrochloride & 1.469 (1.304, 1.660) & 1.429 (1.273, 1.630) & 1.428 (1.265, 1.612) & 1.467 (1.307, 1.648) \\
Paroxetine Mesylate & 1.254 (1.000, 1.718) & Not selected & Not selected & Not selected \\
Trazodone Hydrochloride & 0.608 (0.564, 0.651) & 0.597 (0.545, 0.643) & 0.599 (0.551, 0.646) & 0.611 (0.563, 0.653) \\
Vortioxetine Hydrobromide & 0.862 (0.528, 1.078) & 0.781 (0.485, 1.000) & 0.773 (0.479, 1.000) & 0.823 (0.525, 1.042) \\
Ketamine Hydrochloride & Not selected & 1.326 (1.000, 2.380) & 1.350 (1.000, 2.430) & Not selected \\
Fluoxetine Hydrochloride/Olanzapine & Not selected & Not selected & Not selected & 0.868 (0.599, 1.057) \\
\bottomrule
\end{tabular}
\end{table}

We carefully examined how different approaches to incorporating co-prescription information affected the selection of antidepressant medications. Table~\ref{tab:antidepressant_comprehensive} provides a clear comparative overview, displaying selected antidepressants under tetrachoric correlation, Pearson correlation, conditional probability-based correlation, and the independence scenario. Distinct differences in the subsets of selected antidepressants were observed depending on the correlation method utilized, highlighting the sensitivity of selection results to assumptions about drug co-prescription patterns.  Importantly, this antidepressant-only analysis reflects comparisons within the antidepressant class itself. As a result, fluoxetine hydrochloride appears as increased risk in this restricted analysis but shows decreased risk in the full analysis across all 922 drugs.

Specifically, the baseline independence model identified nine antidepressants associated with both increases and decreases in suicide-related outcomes: \textit{Bupropion Hydrochloride}, \textit{Citalopram Hydrobromide}, \textit{Fluoxetine Hydrochloride}, \textit{Fluoxetine Hydrochloride/Olanzapine}, \textit{Fluvoxamine Maleate}, \textit{Mirtazapine}, \textit{Paroxetine Hydrochloride}, \textit{Trazodone Hydrochloride}, and \textit{Vortioxetine Hydrobromide}. Under the tetrachoric correlation method, which captures latent continuous relationships between drugs, \textit{Paroxetine Mesylate} was uniquely selected, reflecting its strong positive latent correlation (approximately \(+0.45\)) with the closely related antidepressant \textit{Paroxetine Hydrochloride}. This strong latent dependency encouraged borrowing strength, amplifying Paroxetine Mesylate's posterior signal and facilitating its selection. Conversely, Ketamine Hydrochloride and Fluoxetine Hydrochloride/Olanzapine, which lacked strong tetrachoric ties, were excluded under this structure, emphasizing the sharper differentiation of drug-specific effects provided by tetrachoric correlations.

When incorporating the Pearson correlation structure, however, noticeable shifts occurred compared to the selected subset under the independence model: \textit{Ketamine Hydrochloride} entered the selected subset, while \textit{Fluoxetine Hydrochloride}, its combination with Olanzapine, and \textit{Paroxetine Mesylate} were dropped. Examination of the correlation heatmap (Figure~\ref{fig:coprescription_matrix}) reveals that these changes are explained by moderate positive Pearson correlations (approximately \(+0.3\)) between Ketamine and several selective serotonin reuptake inhibitors (SSRIs), effectively increasing Ketamine's posterior signal through borrowing strength across correlated drugs.

Lastly, the conditional probability-based correlation model produced a selection subset closely resembling the independence scenario but omitted \textit{Paroxetine Mesylate}. Inspection of the heatmap indicates relatively low directional conditional probability correlations between Paroxetine Mesylate and other selected antidepressants, providing minimal evidence for shared prescribing patterns, thus explaining its omission.

Collectively, these observations demonstrate that the differences in antidepressant selection across correlation methods are directly interpretable through the co-prescription heatmaps. Correlation structures not only quantify drug-drug dependencies but also substantially impact pharmacovigilance conclusions, highlighting the critical role of carefully choosing a correlation structure aligned with clinical prescribing realities and statistical modeling goals.  Finally, this example also illustrates how class-specific analyses can yield quite different results than analyses that compare all drugs.  The choice depends on the question being asked.

\section{Discussion}

The High Dimensional Empirical Bayes Screening (iDEAS) algorithm is a novel approach to high-dimensional drug safety screening.  \cite{Gibbons2019} turned the problem around by swapping the right and left hand-sides of the model.  Traditional approaches use a single drug to predict a single adverse event using spontaneous reports and computing proportional reporting ratios (\cite{DuMouchel1999}), or using self-controlled case series designs based on claims data (\cite{Farrington1995, Whitaker2006}). \cite{Simpson2013} extended the self-controlled case-series design to exploration of multiple drugs using an approach based on regularized regression, where the drugs enter the equation on the right hand-side of the model.  By contrast, iDEAS places the drugs on the left-hand side of the model and models them jointly by applying random-effects regression to the data arrangement in which  patients are clustered within drugs, which then borrows strength across drugs.  The advantage of this approach is that as the number of drugs increases, so does the number of clusters and precision actually increases (see \cite{Hedeker2006}), whereas the reverse is true for the regularized regression approach.  The disadvantage is that \cite{Gibbons2019} assume an overly simplistic correlational structure across the drugs (i.e., compound symmetry) and use the Bonferroni inequality to adjust for multiplicity.  In this paper we have shown how we can overcome these limitations of the original iDEAS algorithm through use of a Bayesian hierarchical regression model with slab and spike prior distribution to accommodate the high dimensionality and low event probability, quite general correlational structure across the clusters (in our case drugs), and a more principled approach to limiting false discovery.  The net result is that we have been able to identify new drugs that have safety signals (risks and benefits) that are either related to other drugs that we previously identified and therefore biologically plausible, or drugs that are independent of earlier results but have prior evidence of an adverse association (e.g., montelukast and varenicline).  Whether these adverse associations are causal is an open question.  

{

In this paper, we treat the co-prescription matrix $\Sigma_D$ as an external, data-derived input via conditional probability, Pearson correlation, or tetrachoric correlation. A natural next step is a fully Bayesian joint model that places a prior directly on $\Sigma_D$ (or its correlation/precision parameterization) and learns it alongside drug effects. Several principled options exist: priors for correlation matrices such as LKJ (\cite{Lewandowski2009}); separation strategies that model standard deviations and correlations separately (\cite{Barnard2000}); positive-definite parameterizations via Cholesky/log-Cholesky; or sparse graphical formulations using $G$-Wishart priors on precision matrices to encode conditional independences in prescribing networks (\cite{Roverato2002}). The main challenges are computational and inferential. Computationally, joint estimation of a high–dimensional co–prescription covariance is costly: with $p$ drugs, $\Sigma_D$ has $O(p^2)$ free parameters; dense samplers typically require $O(p^3)$ linear algebra per iteration and $O(p^2)$ memory, which is prohibitive for hundreds of drugs and long chains. Inferentially, allowing the outcome to update $\Sigma_D$ can induce feedback—learning a borrowing structure partly driven by the very associations being detected—thereby inflating cluster–specific signals or masking cross–cluster effects. In practice, strong regularization toward independence (e.g., shrinkage/sparsity or graph constraints) is needed to curb spurious correlation–driven borrowing. We view a hierarchically regularized, graph–aware prior for $\Sigma_D$ paired with explicit sensitivity controls and scalable inference as a promising direction for future work.
}

The approach developed here is relevant to any high-dimensional problem with binary outcome and longitudinal exposure data from which pre-versus post exposure event tabulation is possible.  For example, there are many high dimensional problems in molecular genetics that could be analyzed using the modified iDEAS algorithm.

Finally, our approach is not without limitations.  First, in our motivating example, we rely upon ICD codes to determine suicide attempts, which are likely underestimates of the true number of attempts.  This is true for almost any adverse event target to which iDEAS could be applied.  Second, our results are not immune to reverse causality, where it may be the case that the adverse event leads to treatment rather than treatment leading to the adverse event.  This is certainly possible in our example, where for example, a suicide attempt, diagnosis of depression, and prescription of antidepressant can all occur within a very short period of time (even the same day), too short to be a consequence of initiating the drug.


\bibliographystyle{agsm2}
\bibliography{references}

\end{document}